\documentclass[aps, prl, twocolumn, superscriptaddress, nofootinbib, amsfont, graphicx, preprintnumbers]{revtex4-1}

\usepackage{graphicx}
\usepackage{hyperref}
\makeatletter
\@ifpackageloaded{xcolor}{}{\usepackage[table]{xcolor}}
\makeatother
\usepackage{epsfig}
\usepackage{ifpdf}
\usepackage{bm}
\usepackage[english]{babel}
\usepackage{amssymb}
\usepackage{braket}
\usepackage{array}
\usepackage{slashed}
\usepackage{enumerate}
\usepackage{url}
\usepackage{verbatim}
\usepackage{amsthm,amssymb}
\usepackage{mathrsfs}
\usepackage{setspace, gensymb, hhline, booktabs}
\usepackage{tikz-feynman}
\usepackage{amsmath,cleveref} 
\usepackage{graphicx}
\usepackage{tabularx}
\usepackage{multirow}
\usepackage[table]{xcolor}
\usepackage[normalem]{ulem}

\usepackage[absolute,overlay]{textpos} 
\textblockorigin{0cm}{0cm}
\begin{document}
%-----
% \begin{textblock}{6}(14,1)
%   \raggedleft
%   {\text{CPTNP-2025-005}}  
% \end{textblock}
\preprint{CPTNP-2025-029}
%-----

% \title{Search for Axions in REactor-based COherent Dark matter Experiment}
\title{Hunting for Axions in REactor neutrino COherent scattering Detection Experiment}
\author{Wei Dai}
\altaffiliation{These authors contribute equally}
\affiliation{Department of Physics and Institute of Theoretical Physics, Nanjing Normal University, Nanjing, 210023, China}
% \affiliation{Nanjing Key Laboratory of Particle Physics and Astrophysics, Nanjing, 210023, China}
\author{Yuanlin Gong}
\altaffiliation{These authors contribute equally}
\affiliation{Department of Physics and Institute of Theoretical Physics, Nanjing Normal University, Nanjing, 210023, China}
% \affiliation{Nanjing Key Laboratory of Particle Physics and Astrophysics, Nanjing, 210023, China}
\author{Guanhua Gu}
\altaffiliation{These authors contribute equally}
\affiliation{Institute of Theoretical Physics, Chinese Academy of Sciences}
\author{Liangliang Su}
\altaffiliation{These authors contribute equally}
\affiliation{Department of Physics and Institute of Theoretical Physics, Nanjing Normal University, Nanjing, 210023, China}
% \affiliation{Nanjing Key Laboratory of Particle Physics and Astrophysics, Nanjing, 210023, China}
\author{Li Wang}
\email{Corresponding author: wangl@bnu.edu.cn}
\affiliation{School of Physics and Astronomy, Beijing Normal University, Beijing 100875}
\author{Lei Wu}
\email{Corresponding author: leiwu@njnu.edu.cn}
\affiliation{Department of Physics and Institute of Theoretical Physics, Nanjing Normal University, Nanjing, 210023, China}
\affiliation{Nanjing Key Laboratory of Particle Physics and Astrophysics, Nanjing, 210023, China}
\author{Yongcheng Wu}
\email{Corresponding author: ycwu@njnu.edu.cn}
\affiliation{Department of Physics and Institute of Theoretical Physics, Nanjing Normal University, Nanjing, 210023, China}
\affiliation{Nanjing Key Laboratory of Particle Physics and Astrophysics, Nanjing, 210023, China}
\author{Litao Yang}
\email{Corresponding author: yanglt@mail.tsinghua.edu.cn}
\affiliation{Key Laboratory of Particle and Radiation Imaging (Ministry of Education) and Department of Engineering Physics, Tsinghua University, Beijing 100084}

\date{\today}% It is always \today, today,

\begin{abstract}
Nuclear power plants are not only vital sources of clean energy but also powerful facilities for probing new physics beyond the Standard Model. Due to the intense gamma-ray flux and an appropriate energy condition, they are particularly well-suited for searches of light hypothetical particles such as sub-MeV axions and axion-like particles (ALPs). In this work, we present a sensitivity study for ALPs searches with the REactor Neutrino COherent scattering Detection Experiment (RECODE), where two low-threshold, high-purity germanium detectors are placed at 11 m (near point) and 22 m (far point) from a 3.4 GW nuclear reactor at Sanmen nuclear power plant. With a 10 kg$\cdot$year exposure, we demonstrate that the expected sensitivities to the ALP couplings to electrons and photons are competitive with or surpass the available results from  beam-dump experiments. A planned upgrade to 100 kg$\cdot$year will {nearly cover} the so-called {\it cosmological triangle} region, probing unexplored parameter space relevant to axions.

\end{abstract}

\pacs{Valid PACS appear here}% PACS, the Physics and Astronomy
\maketitle

{\it \textbf {Introduction}} --
The Peccei-Quinn (PQ) mechanism~\cite{Peccei:1977hh} provides an elegant solution to the Strong CP Problem in Quantum Chromodynamics (QCD) by introducing a global chiral U(1) symmetry. The spontaneous breaking of this symmetry generates a pseudo-Goldstone boson, the axion~\cite{Weinberg:1977ma,Wilczek:1977pj}. The original QCD axion was soon ruled out by experiments due to its sizable couplings. Then, the invisible axion models were proposed~\cite{Kim:1979if,Shifman:1979if,Zhitnitsky:1980tq,Dine:1981rt}. Besides, many UV-complete theories like string theory~\cite{Witten:1984dg} naturally predict the axion-like particles~\footnote{In this paper, we use ALPs to refer to axions or axion-like particles.}, which would not be related to the strong CP problem. The couplings and masses of these axions do not follow the QCD axion relation, and can span many orders of magnitude~\cite{Preskill1983,Gaitskell:2004gd,Svrcek:2006yi,Brivio:2017ije}. 

% \textcolor{red}{Accordingly, ALPs have also been widely discussed in cosmological and astrophysical scenarios, including pulsar timing array observations, high-redshift structures, and primordial structure formation~\cite{Xie:2022uvp,Guo:2023hyp,Rubin:2024scpma}.}

Significant efforts have been made in searching for low-mass ALPs, covering helioscope, haloscope and their variants, like ADMX \cite{ADMX:2018gho,ADMX:2019uok}, CAST \cite{CAST:2004gzq,CAST:2007jps,CAST:2017uph}, ABRA \cite{Ouellet:2018beu,Salemi:2021gck}, CASPEr \cite{Budker:2013hfa,JacksonKimball:2017elr}, HAYSTAC \cite{HAYSTAC:2018rwy}, and dark matter direct detection experiments \cite{PandaX:2017ock,LUX:2017glr,SuperCDMS:2019jxx}(see refs. \cite{Irastorza:2018dyq,Sikivie:2020zpn} for recent reviews). Meanwhile, the model-building works have renewed interest in heavy ALPs within the MeV-to-GeV mass range~\cite{Hook:2014cda,Fukuda:2015ana,Dimopoulos:2016lvn,Hook:2019qoh}, which may address the PQ quality problem~\cite{Kamionkowski1992,Holman1992,Ghigna1992,Barr1992,Hook:2019qoh}. For such ALPs, the collection of the astrophysical bounds~\cite{Giannotti:2015kwo,Janka:2017vcp,Bollig:2020xdr,Lucente:2020whw,Ferreira:2022xlw,Lucente:2022wai} and terrestrial constraints~\cite{Riordan:1987aw,Mimasu:2014nea,Jaeckel:2015jla,Dobrich:2015jyk,Bauer:2017nlg,Bauer:2017ris,Dolan:2017osp,Harland-Lang:2019zur,Gavela:2019cmq,Wang:2021uyb,Ren:2021prq,CCM:2021jmk,Capozzi:2023ffu} leaves an uncharted region in the parameter space, i.e. the {\it cosmological triangle}, for ALP masses of $0.3\ \mathrm{MeV} \lesssim m_a \lesssim\  0.9\  \mathrm{MeV}$ and the ALP-photon coupling of $1.3\times 10^{-5}\ \mathrm{GeV^{-1}}\lesssim g_{a\gamma}\lesssim\ 9.1\times 10^{-5}\ \mathrm{GeV^{-1}}$~\cite{Dent:2019ueq,Brdar:2020dpr}. Supernova cooling arguments can constrain this parameter space ~\cite{Caputo:2021rux}, but depend on the assumptions regarding the muonic supernova core and the adopted Garching supernova model ~\cite{Janka:2017vcp,Bollig:2020xdr}. Although model-dependent cosmological constraints from Big Bang Nucleosynthesis and the Cosmic Microwave Background can also exclude it ~\cite{Depta:2020zbh,Balazs:2022tjl}, they can be evaded in non-standard cosmological scenarios ~\cite{Carenza:2020zil,Depta:2020wmr}.

\begin{figure}
    \centering
    \includegraphics[width=1.0\columnwidth]{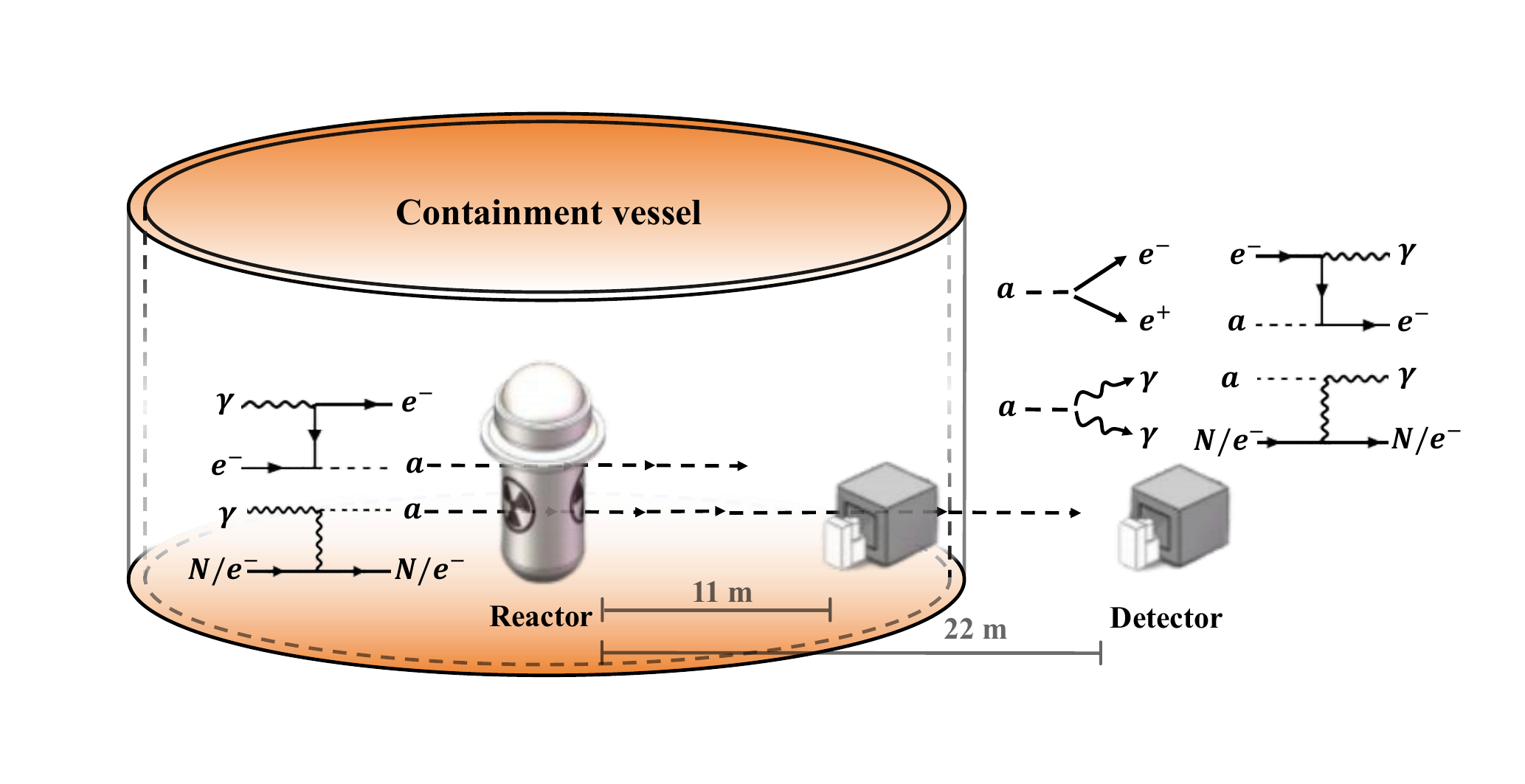}
        \caption{
         “Schematic of ALPs production, decay, and detection. RECODE utilizes
         % Cartoon of the ALPs production, decay, and detection in the RECODE. 
        % In nuclear reactors, ALPs are produced via the Primakoff effect and Compton scattering. During propagation, they may decay, and upon reaching the detector, their energy deposition can be measured through inverse Primakoff scattering, inverse Compton scattering and decay.
       % The experiment utilizes 
        a dual-detector configuration, comprising a near-point detector positioned inside the containment vessel at a distance of 11 m from the reactor core and a far-point detector located outside the containment at 22 m from the core. The detectors contain multi-layer shielding assembly (Cu/Pb/polyethylene) specifically designed to suppress cosmic-ray neutrons and background radiation.
        % The containment consists of a multilayer shielding assembly (Cu/Pb/polyethylene) specifically designed to suppress cosmic-ray neutrons and background radiation.
        }
        \label{fig:RECODE}
\end{figure}
Leveraging intense particle fluxes and favorable energy conditions, nuclear reactors provide a critical experimental platform for studying neutrinos, dark matter, and other new physics phenomena~\cite{FernandezMoroni:2014qlq,Belov:2015ufh,Akimov:2017hee,Arias-Aragon:2023ehh,TEXONO:2018nir,RELICS:2024opj,CONNIE:2024off}. Notably, they can serve as highly efficient facilities for ALP production. Previous experimental searches based on nuclear reactors have explored sensitivity to MeV-scale ALPs, including the interactions such as axion-neutron~\cite{Cavaignac:1982ek,JacksonKimball:2017elr,NEON:2024kwv}, axion-photon~\cite{Dent:2019ueq,AristizabalSierra:2020rom,ADMX:2021nhd,BREAD:2021tpx,NEON:2024kwv,Sahoo:2024zee,Mirzakhani:2025bqz}, and axion-electron~\cite{PandaX:2017ock,Dent:2019ueq,AristizabalSierra:2020rom,SuperCDMS:2019jxx,NEON:2024kwv}, as well as their combinations~\cite{TEXONO:2006spf}.
%coherent elastic neutrino-nucleus scattering (CE$\nu$NS)\cite{CONNIE:2019swq,MINER:2016igy,Hakenmuller:2019ecb}, which offer a robust platform for conducting comprehensive studies on axions, thereby advancing our understanding of their properties and interactions.

In this study, we investigate the sensitivity of the ongoing REactor neutrino COherent scattering Detection Experiment (RECODE)~\cite{Yang:2024exl} to probe the ALP-photon and ALP-electron couplings. RECODE is currently on building at Sanmen nuclear power plant. Our  analysis will be performed according to the projected detector
exposure and background simulation. The ALPs could be produced via Primakoff~\cite{Wong:2015kgl} and Compton-like processes in the reactor, and then are detected through their inverse processes or decay processes in the detectors, as illustrated in Fig.~\ref{fig:RECODE}. RECODE comprises two arrays of low-threshold germanium detectors, each with a total mass of 5 kg. {Similar low-threshold germanium-detector technologies have been extensively developed in rare-event searches, such as CDEX\cite{Ma:2018tdv,Jiang:2018lij,CDEX:2020tkb}.} These arrays are positioned at distances of 11 m and 22 m from the reactor core of the Sanmen Nuclear Power Plant. The implementation of a near-far detector design can significantly reduce the background and systematic uncertainties. Due to its high thermal power reactor source and short baseline configuration (see Table. \ref{tab:reactor_experiments} for the comparison), it has a great potential to probe previously inaccessible regions of parameter space. In its ongoing first run, RECODE will achieve an exposure of 10 kg$\cdot$year with an anticipated background of 2 \(\mathrm{keV^{-1} kg^{-1} day^{-1}}\)(dru). With these parameters, RECODE would surpass the sensitivity of existing beam-dump experiments in constraining the axion-photon coupling $g_{a\gamma}$ in the cosmological triangle region. For a future upgrade with an exposure of 100 kg$\cdot$year and 0.2 dru background, RECODE {can test} the cosmological triangle and produce the competitive bounds on the ALP-electron coupling in comparison with other experiments. These results will be world-leading in the mass range of 0.3 MeV $ < m_a < 1$ MeV and establish new benchmarks for ALPs searches.

{\it \textbf{Signals of ALPs} } -- We consider the following interactions of the ALP field $a$ with the photon field $A_\mu$ and electron field $\psi_e$,
\begin{equation}
\mathcal{L}_{\mathrm{int}} = -\frac{1}{4}g_{a\gamma}aF_{\mu\nu}\tilde{F}^{\mu\nu} - g_{ae}a \bar{\psi}_{e}\gamma_{5}\psi_{e},
\end{equation}
where $F_{\mu\nu}$ and $\tilde{F}_{\mu\nu}$ are the electromagnetic field strength tensor and its dual, respectively. In our study, we assume only one coupling involved in the relevant processes. In the presence of the axion-photon coupling, ALPs can be produced via Primakoff scattering in the reactor, where the incident photons interact with nucleon targets $\gamma + N \to a + N$. The differential cross section for this process is given by
\begin{equation}
	\begin{split}
		\frac{{d}\sigma_{{P}}}{dt} =  & \frac{e ^{2} Z^{2}F^{2}(t)\,g_{a\gamma}^{2}M_{N}^{2}}{8 \pi t^{2} E_{\gamma}^{2} (t-4M_{N}^{2})^{2}}\bigg\{2 m_{a}^{2} t (M_{N}^{2}+E_{\gamma}M_{N}) - \\&m_{a}^{4} M_{N}^{2} -t\left[4E_{\gamma}^{2} M_{N}^{2} +t(M_{N}^{2}+2E_{\gamma}M_{N})\right]\bigg\},
	\end{split}
\end{equation}
% \begin{equation}
% 	\begin{split}
% 		\frac{{d}\sigma_{{P}}}{{d}t} = & \, 2\alpha Z^{2}F^{2}(t)\,g_{a\gamma}^{2}\frac{M_{N}^{4}}{t^{2}(M_{N}^{2}-s)^{2}(t-4M_{N}^{2})^{2}} \\
% 		& \times \left\{m_{a}^{2}t(M_{N}^{2}+s)-m_{a}^{4}M_{N}^{2}-t\left[(M_{N}^{2}-s)^{2}+st\right]\right\},
% 	\end{split}
% \end{equation}
where $t$ is the Mandelstam variable and $Z$ is the atomic number of the target material, $^{235}U$. $F(t) = a_0^2 t/(1-a_0^2t)$ is the atomic nucleus form factor with $a_0=111.74/Z^{1/3}m_e$. $E_\gamma$ is the energy of the incident photon and $M_N$ is the nucleon mass. 
Since the nucleon mass is much larger than the incident photon energy, the energy $E_a$ of the produced ALPs nearly equals the energy of the incident photon, $E_a \approx E_\gamma$. Using the 4-momentum conservation and the photon on-shell condition, we obtain the kinematic boundaries for the process
\begin{equation}\label{eq:kenetic_photon}
    t_{\min(\max)}=\frac{m_a^4}{4s} - \left(p_{\gamma \text{cm}} \pm k_{a \text{cm}}\right)^2,
\end{equation}
where $p_{\gamma\text{cm}}$ and $k_{a \text{cm}}$ denote the incident photon and outgoing ALP three-momenta, respectively, in the center-of-momentum frame \cite{AristizabalSierra:2020rom}.

% \begin{table}[!bp]
% \centering
% \caption{Comparison of various reactor neutrino experiments on the thermal power and baseline from the reactor to detector. As the ALP flux at the detector is proportional to the ratio of $P/L^2$, the last column shows its values for different experiments.}
% %Comparison of reactor neutrino experiments by power-to-baseline ratio
% \label{tab:reactor_experiments}
% \begin{tabularx}{\columnwidth}{|>{\raggedright\arraybackslash}X|>{\centering\arraybackslash}p{1.8cm}|>{\centering\arraybackslash}p{1.8cm}|>{\centering\arraybackslash}p{1.8cm}|}
% \hline
% \multirow{2}{*}{\textbf{Experiment}} & \textbf{Power  ($P$/GW)} & \textbf{Baseline ($L$/m)} & \textbf{$P/L^2$ (GW/m$^2$)} \\
% \hline
% RECODE & 3.4 & 11 & 0.0281 \\ 
% NEON ~\cite{NEON:2024kwv} & 2.8 & 23.7 & 0.0050 \\
% KNPP ~\cite{Belov:2015ufh} & 1 & 11.2 & 0.0080 \\
% TEXONO ~\cite{Wong:2015kgl} & 2.9 & 28 & 0.0037 \\
% $\nu$CLEUS ~\cite{NUCLEUS:2017htt} & 4 & 15 & 0.0178 \\ 
% RED-100 ~\cite{Akimov:2017hee} & 1 & 19 & 0.0028 \\ 
% CONNIE ~\cite{FernandezMoroni:2014qlq} & 3.95 & 30 & 0.0044 \\ 
% vIOLETA ~\cite{Fernandez-Moroni:2020yyl} & 2 & 12 & 0.0139 \\ 
% SoLid ~\cite{SoLid:2020cen} & 0.065 & 6.3 & 0.0016 \\ 
% CONUS+ ~\cite{CONUS:2024lnu} & 3.6 & 20.7 & 0.0084 \\ 
% $\nu$GeN ~\cite{nGeN:2022uje} & 3.1 & 11.8 & 0.0222 \\ \hline
% \end{tabularx}
% \end{table}
\begin{table}[!bp]
\centering
\caption{Comparison of various reactor neutrino experiments in terms of }the thermal power and baseline from the reactor to detector. As the ALP flux at the detector is proportional to the ratio of $P/L^2$, the last column shows its values for different experiments.
\label{tab:reactor_experiments}
\begin{ruledtabular}
\begin{tabular}{lccc}
\shortstack{\textbf{Experiment}\\\textbf{}}
& \textbf{\shortstack{Power\\($P$/GW)}} 
& \textbf{\shortstack{Baseline\\($L$/m)}} 
& \textbf{\shortstack{$P/L^2$\\(GW/m$^2$)}}\\
RECODE & 3.4 & 11 & 0.0281 \\ 
NEON~\cite{NEON:2024kwv} & 2.8 & 23.7 & 0.0050 \\
KNPP~\cite{Belov:2015ufh} & 1 & 11.2 & 0.0080 \\
TEXONO~\cite{Wong:2015kgl} & 2.9 & 28 & 0.0037 \\
$\nu$CLEUS~\cite{NUCLEUS:2017htt} & 4 & 15 & 0.0178 \\ 
RED-100~\cite{Akimov:2017hee} & 1 & 19 & 0.0028 \\ 
CONNIE~\cite{FernandezMoroni:2014qlq} & 3.95 & 30 & 0.0044 \\ 
vIOLETA~\cite{Fernandez-Moroni:2020yyl} & 2 & 12 & 0.0139 \\ 
SoLid~\cite{SoLid:2020cen} & 0.065 & 6.3 & 0.0016 \\ 
CONUS+~\cite{CONUS:2024lnu} & 3.6 & 20.7 & 0.0084 \\ 
$\nu$GeN~\cite{nGeN:2022uje} & 3.1 & 11.8 & 0.0222 \\
\end{tabular}
\end{ruledtabular}
\end{table}
The generated ALPs will flight out the reactor and the outgoing ALPs flux at the detector is obtained by convolving the incident photon flux with the normalized differential cross section, which can be expressed as
\begin{equation}\label{eq:flux}
  \frac{d\Phi_{a}}{dE_{a}}=P_{\mathrm{sur}}\int _{E_{\gamma}^\mathrm{min}}^{E_{\gamma}^\mathrm{max}}\frac{1}{\sigma_{\mathrm{tot}}}\frac{d\sigma_{P}}{dE_{a}}(E_{\gamma},E_a)\frac{d\Phi_{\gamma}}{dE_{\gamma}}dE_{\gamma},
\end{equation}  
where $E^{\mathrm{min}}_{\gamma}$ and $E^{\mathrm{max}}_{\gamma}$ is the kinetic boundary determined by Eq.~(\ref{eq:kenetic_photon}), and the total cross section is given by $\sigma_{\mathrm{tot}} = \sigma_{\mathrm{SM}} + \sigma_{P}$.
% For the small axion-photon coupling $g_{a\gamma}$ considered here, the Primakoff contribution $\sigma_{P}$ is negligible compared to $\sigma_{\mathrm{SM}}$,
% leading to $\sigma_{\mathrm{Tot}}\approx\sigma_{\mathrm{SM}}$.
Here we utilize the Standard Model cross section $\sigma_{\mathrm{SM}}$ for total photon scattering cross-section in the reactor, which can be directly obtained from the database~\cite{XCOM}.  For the reactor photon flux, we employ the well-established analytical approximation derived by~\cite{bechteler_faissner_yogeshwar_seyfarth_1984}
\begin{equation}
  \frac{d\Phi_{\gamma}}{dE_{\gamma}}=\frac{5.8\times 10^{17}}{\mathrm{MeV\cdot sec}}\left(\frac{P}{\mathrm{MW}}\right)e^{-1.1E_{\gamma}/\mathrm{MeV}},
\end{equation}
where the thermal power is $P = 3400\, \rm{MW}$ for RECODE. This model holds for photon energies above 0.2 MeV. The survival probability of the ALP during its propagation is evaluated by ${P_{\rm sur}\equiv \exp(-L{m_a}\Gamma_a/|\vec{p}_a|)}$, where the decay width to photons is $\Gamma_{2\gamma} = g_{a\gamma}^2 m_a^3/64\pi$ and $L$ is the distance from the reactor to the detector.

Inside the germanium detector, the inverse Primakoff process offers a distinctive signature, characterized by a monoenergetic photon with energy $E_{\gamma^{\prime}}\approx E_a$. At higher ALPs masses, in-flight decays of ALPs become significant and produce two photons in the final states. Including both processes, we can obtain the signal event rate induced by the ALP-photon coupling as
\begin{align}
 \frac{dN}{dE_a} = \left( \frac{1}{4\pi L^2} \frac{d\Phi_a}{d{E_a}} \right)  \left( m_{\mathrm{det}} {N_T} \sigma_D 
+ {S} P_{\text{decay}}\right) \cdot \Delta t ,
\label{eq:6}
\end{align}	
where $N_T = 8.29 \times 10^{24}$ $\mathrm{kg}^{-1}$ is the target nuclear surface density in the detector, $\sigma_D$ is the scattering cross section for the inverse Primakoff process, $S = L_{\mathrm{det}}^2=0.106 \times 0.106$ $\mathrm{m}^2$ is the detector transverse area, $m_{\mathrm{det}} = 5$ kg is the mass of each detector, and $\Delta t = 1$ year stands for the data-taking time. The probability for an ALP to decay within the detector is given by {$P_{\rm decay} = P_{\rm sur} (L) - P_{\rm sur}(L+L_{\rm det})$}, with $L_{\mathrm{det}}$ being the detector length. 
%We note that Meanwhile, when particles enter the detector, electron and photon kinetic energy can convert to photon energy and be collected, other particle kinetic energy present a ratio during this conversion, called quenching factor\cite{Demirci:2024vzk}. We can neglect this factor for our photon subject and electron followed.  

\begin{figure}
    \centering
    \includegraphics[width=\columnwidth]{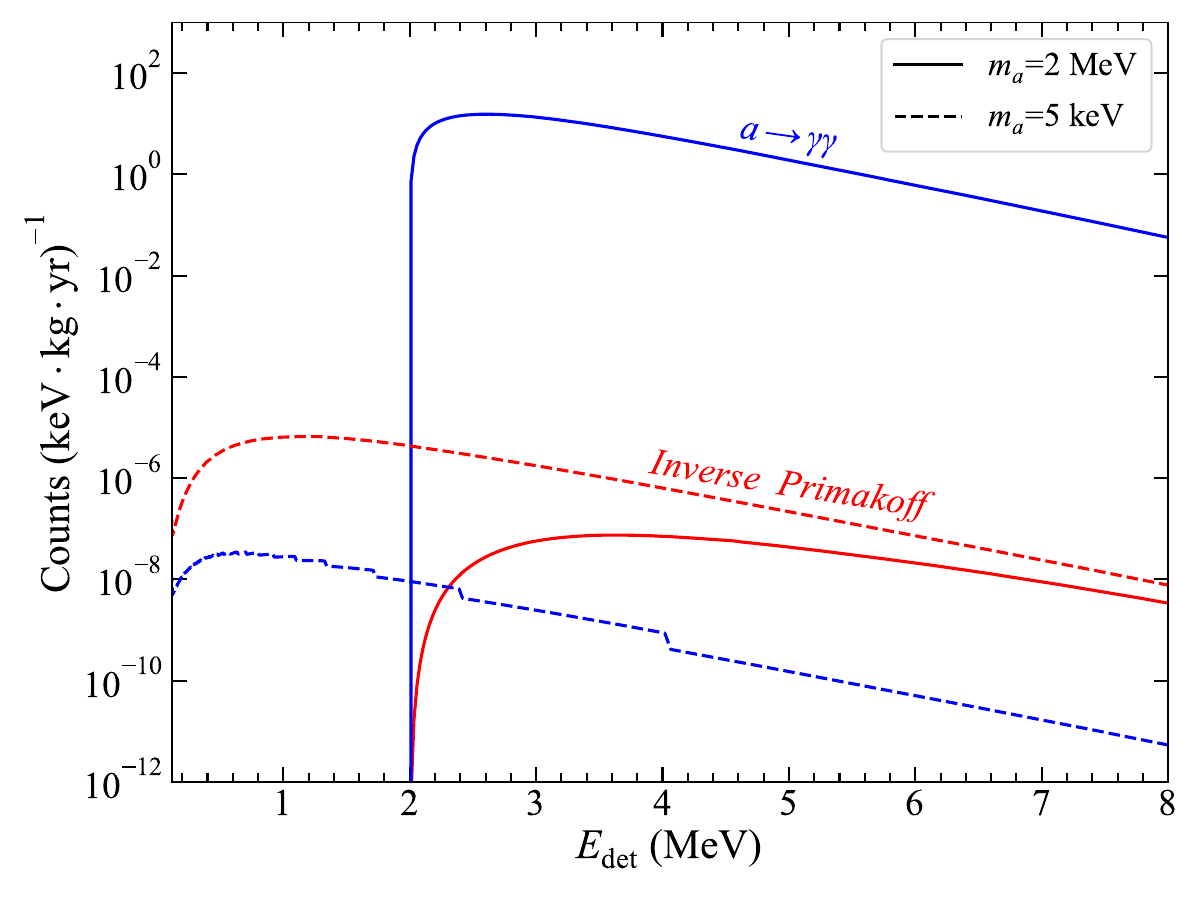}
        \caption{Comparison of event rates between the inverse Primakoff process (red lines) and diphoton decay $a\to \gamma\gamma$ (blue lines) for a light (dashed lines) and heavy (solid lines) ALP}. The coupling is fixed at  $g_{a\gamma} = 2 \times 10^{-5} ~\mathrm{GeV^{-1}}$ and $g_{ae} = 0$.
    \label{account_plot_gagamma}
\end{figure}
\begin{figure}
    \centering
    \includegraphics[width=\columnwidth]{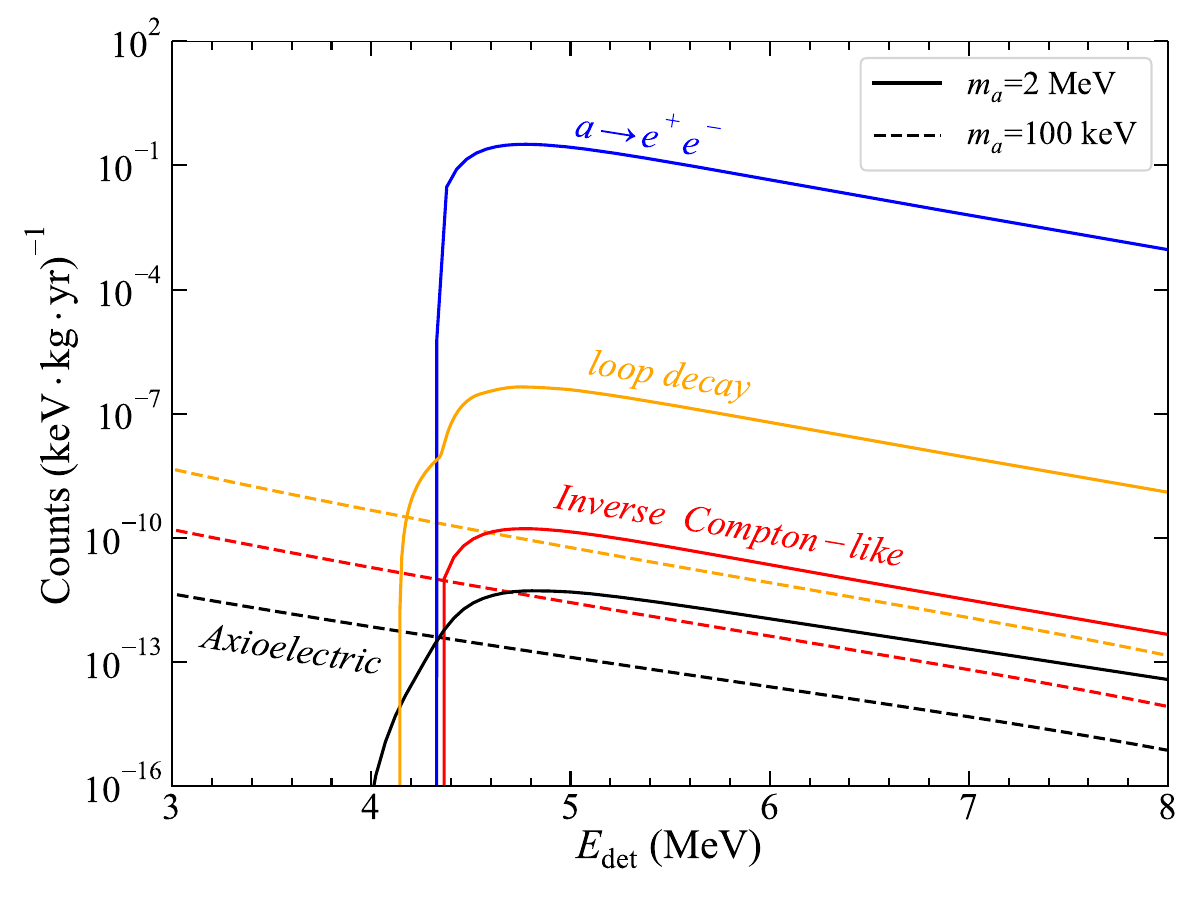}
    %\caption{Same as Fig.~\ref{account_plot_gagamma}, but for the inverse Compton-like process (red lines), the dielectron decay $a\to e^+ e^-$ (blue lines), axioelectric effect (black lines) and the diphoton decay induced by the electron-loop (yellow lines).
    \caption{Comparison of event rates between the inverse Compton-like process (red lines), the dielectron decay $a\to e^+ e^-$ (blue lines), axioelectric effect (black lines) and the diphoton decay induced by the electron-loop (yellow lines) for a light (dashed lines) and heavy (solid lines) ALP.} The coupling is fixed at $g_{ae} = 1 \times 10^{-7}$ and $g_{a\gamma} = 0$.
    \label{account_plot_gae}
\end{figure} 

For the ALP-electron coupling, the flux can be obtained by replacing the differential Primakoff cross section $d\sigma_{P}/dE_a$ in Eq.~(\ref{eq:flux}) with the differential cross section of the Compton-like process $\gamma + e \to a + e$,
% \begin{align}
%   \frac{d\sigma_{C}}{dE_{{a}}}&=\frac{Z\pi g_{ae}^{2}\alpha x}{4\pi (s-m_{e}^{2})(1-x)E_{\gamma}}\left[x-\frac{2m_{a}^{2}s}{(s-m_{e}^{2})}\right. \notag \\
%   &+\left.\frac{2m_{a}^{2}}{(s-m_{e}^{2})}\left(\frac{m_{e}^{2}}{1-x}+\frac{m_{a}^{2}}{x}\right)\right],
% \end{align}
\begin{align}
  \frac{d\sigma_{C}}{dE_{{a}}}&=\frac{Z e^{2} g_{ae}^{2} x}{32 \pi E_{\gamma}^{2} E_{e} (x-1)}\left[x-\frac{m_{a}^{2}(m_e^{2}+2E_{\gamma} E_{e})}{E_{\gamma} E_{e}}\right. \notag \\
  & +\left.\frac{m_{a}^{2}}{E_{\gamma} E_{e}}\left(\frac{m_{e}^{2}}{1-x}+\frac{m_{a}^{2}}{x}\right)\right],
\end{align}
where $E_e$ is the energy of the incident electron, $m_{e}$ is the electron mass and $x=1-E_a/E_{\gamma}+m_{a}^{2}/2E_{\gamma}m_{e}$. Accordingly, the dimensionless variable $x$ is bounded by
\begin{equation}
    x_{\min(\max)}=\frac{E_a\mp k_a}{m_e}.
\end{equation}
When $m_a \ge 2m_e$, the decay channel $a \to e^+e^-$ will be opened and dominate over the inverse Compton-like process. The decay width is given by $  \Gamma_{2e} = {g_{ae}^{2}m_{a}}\sqrt{1-4{m_{e}^{2}}/{m_{a}^{2}}}/{(8\pi)}$. Consequently, we can obtain the signal event rate induced by the ALP-electron coupling with Eq.~(\ref{eq:6}) by using the corresponding cross section, flux and decay probability. Note that when $m_a < 2m_e$, the ALP can also decay to the diphoton via the electron-loop. However, the contribution of such a process is model-dependent. If not suppressed, this decay will be the dominant channel in the low mass range of the ALPs (see the appendix for details).
%For a ALP with mass $m_a < 2m_e$, we consider the decay process $ a \to 2\gamma$ mediated by an electron loop and calculate the decay width as\cite{Bauer:2017ris}
% \begin{equation}
%     \Gamma(a \to  2\gamma)=\frac{g^2_{ae}e^4}{256\pi^5}\frac{m_a^3}{m_e^2}f_p(\frac{m_a^2}{4m_e^2}),
% \end{equation}
% where $f_p(x)=\frac{1}{64x^2} \left|\ln \left[1-2\left(x+\sqrt{x^2-x}\right)\right]^2\right|^2$. When $m_a \ge 2m_e$, the decay channel $a \to e^+e^-$ will be open and its decay width is given by $  \Gamma_{2e} = {g_{ae}^{2}m_{a}}\sqrt{1-4{m_{e}^{2}}/{m_{a}^{2}}}/{(8\pi)}$. Consequently, we can obtain the total event rate with Eq.~(\ref{eq:account}) by using the corresponding detecting cross section, flux and decay probability.

At the detector level, we simulate the energy resolution of the detectors by assuming a Gaussian smearing for the final states, $\sigma\left(E\right)=(35.8+16.6\sqrt{E})$ eV~\cite{CDEX:2016tve,CDEX:2024bum}. Then, the event rate of Eq.~\ref{eq:6} becomes, 
\begin{equation}
    \frac{dN}{dE_{\mathrm{det}}} = \int dE_a \frac{dN}{dE_a}\frac{1}{\sqrt{2\pi}\sigma} e^{-\frac{(E_{\mathrm{det}}-E_a)^2}{2\sigma^2} }.
\end{equation}
In Fig. \ref{account_plot_gagamma}, we compare the event rates of two channels induced by ALP-photon coupling at $m_a=5$ keV and 2 MeV. We can see that the inverse Primakoff process is dominant for the light ALP, while the diphoton decay channel of ALP becomes significant for the heavy ALP. Similarly, we show the results for ALP-electron interactions in Fig.~\ref{account_plot_gae}. We find that the electron-loop mediated decay channel is dominant when $m_a < 2m_e$. Once the dielectron decay channel is allowed kinematically, it will dominate over other channels. In particular, the sharp suppression at lower mass for $a\to e^+e^-$ is because of the high threshold on $E_a$ due to the kinematics constraint from Compton-like scattering during production.

{\it \textbf{\textbf{Experimental sensitivity}}} -- The dominant Standard Model (SM) continuous backgrounds arise from cosmic-ray-induced muons and neutrons which will be suppressed by a multi-layered shield consisting of copper (Cu), lead (Pb) and polyethylene (PE), materials chosen for their dual capability to attenuate both cosmic rays and reactor-generated neutrons. Since RECODE is operated at a surface reactor site without the thick-rock overburden available in underground laboratories, the detector shielding is strengthened to suppress the dominant environmental backgrounds. Further background reduction is achieved through active veto systems: a plastic scintillator veto detector to tag muons, and an NaI scintillator veto detector to identify natural radioactivity gamma rays and cosmic-ray-induced secondary particles. Dedicated on-site background measurements have been performed at the reactor site to characterize the local environmental backgrounds under the present testing conditions, which provide the basis for our background evaluation. Based on these onsite measurements, we further carry out background simulations including the multi-layer shielding and active veto systems described above, to estimate the residual background level inside the detector. After shielding and anti-coincidence veto, the remaining background is dominated by intrinsic radioactivity from the germanium detector structure material, particularly from $^{40}$K decays. Furthermore, for the near detector located
inside the containment vessel, the reactor containment itself can provide
additional shielding against external backgrounds. Combining the onsite measurements, the background simulations including the shielding and veto configuration, and the demonstrated background performance of the same type of low-threshold germanium detectors at CDEX as a realistic reference for the intrinsic detector background, we expect that the SM backgrounds can be controlled to the level of $\mathcal{O}(10)$ dru or even lower level. Therefore, a benchmark background level of
2 dru is expected to be achievable in a large range of energy up to several hundreds keV, while lower level is expected above 1 MeV from the reported data of NEON~\cite{NEON:2024kwv}.

% The dominant Standard Model (SM) continuous backgrounds arise from cosmic-ray-induced muons and neutrons which will be suprressed by a multi-layered shield consisting of copper (Cu), lead (Pb) and polyethylene (PE), materials chosen for their dual capability to attenuate both cosmic rays and reactor-generated neutrons. Further background reduction is achieved through active veto systems: a plastic scintillator veto detector to tag muons, and an NaI scintillator veto detector to identify natural radioactivity gamma rays and cosmic-ray-induced secondary particles. After shielding and anti-coincidence veto, the remaining background is dominated by intrinsic radioactivity from the germanium detector structure material, particularly from $^{40}$K decays. \textcolor{red}{Based on dedicated onsite background measurements and background simulations at the reactor site, together with the demonstrated background performance of the same type of low-threshold germanium detectors at CDEX as a realistic reference for the intrinsic detector background, we expect that the SM backgrounds can be controlled to the level of approximately 2 dru.} Especially, our pre-experiments have shown that a 2 \textcolor{red}{dru} background level is achievable in a large range of energy up to several hundreds keV, while lower level is expected above 1 MeV from the reported data of NEON \cite{NEON:2024kwv}.

\begin{figure}
% [htbp]
    \centering
    \includegraphics[width=\columnwidth]{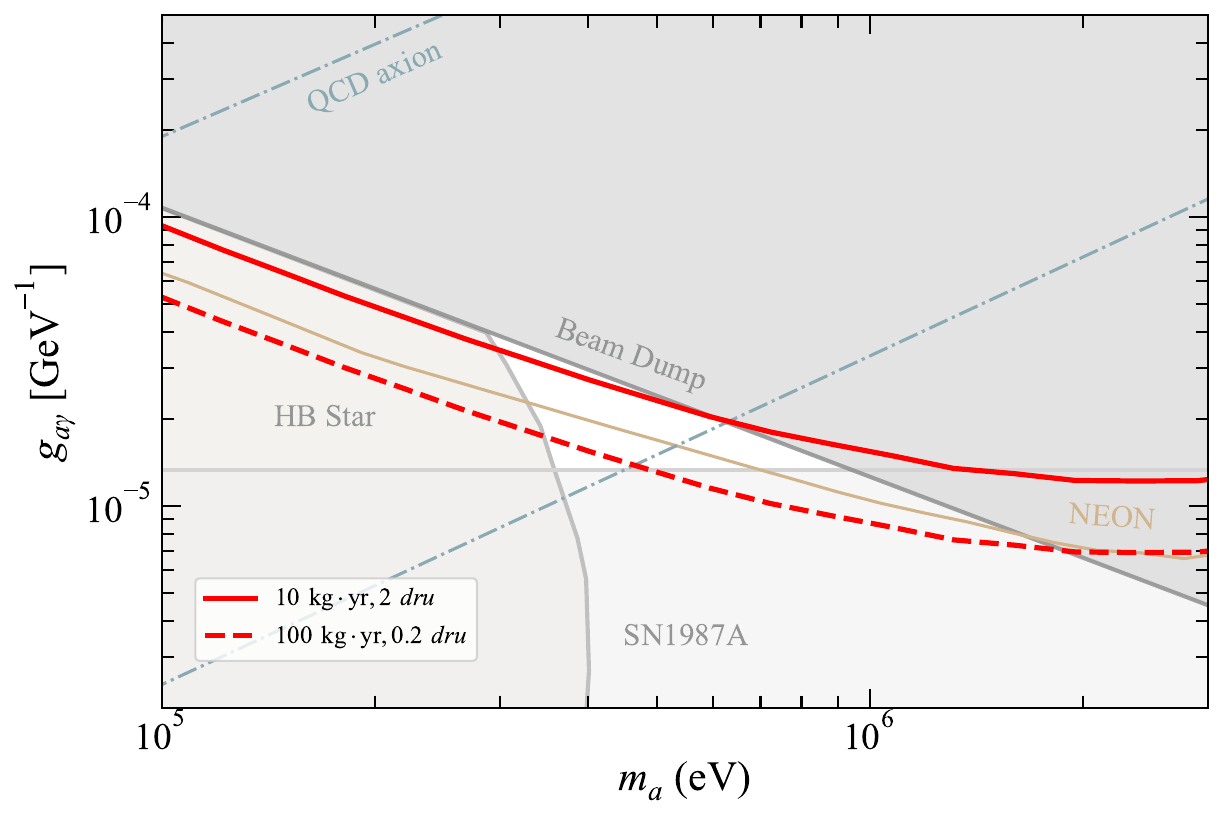}
    \includegraphics[width=\columnwidth]{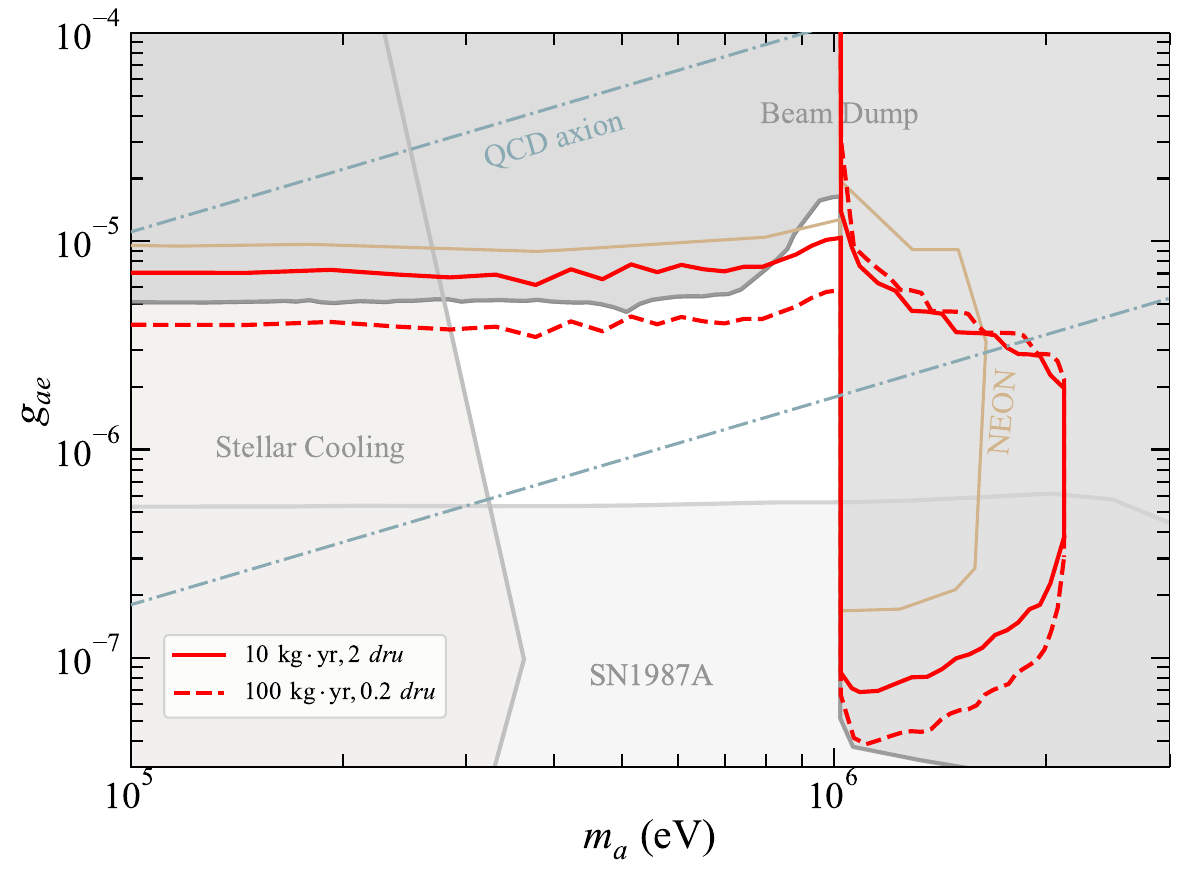}
\caption{The projected 90\% C.L. exclusion limits for the ALP-photon coupling (top panel) and ALP-electron coupling (bottom panel). The constraints from  HB stars cooling~\cite{Carenza:2020zil}, SN1987A~\cite{Lucente:2020whw,Ferreira:2022xlw}, Beam Dump experiments~\cite{Bjorken:1988as,Bechis:1979kp,1987Search,Blumlein:1991xh,Andreas:2010ms,CCM:2021jmk,Jaeckel:2015jla,Dobrich:2015jyk}, stellar cooling~\cite{Hardy:2016kme}, and NEON~\cite{NEON:2024kwv} are included. The parameter space encompassed by QCD axion~\cite{DiLuzio:2020wdo} is also presented.}
    \label{fig:tree}
\end{figure}

The sensitivity of RECODE is estimated by constructing $\chi^2$ from seven 700-keV bins from 100 keV to 5000 keV. {Both statistical and systematic uncertainties are included. The systematic uncertainty, taken to be 10\%, accounts for the reactor power uncertainty, background flux uncertainty, and the overall detector efficiency uncertainty.} The systematics depends on the realistic detector performances. {We also assume 1:1 reactor on-off ratio to reduce the backgrounds. We anticipate our analysis can be improved by optimizing signal extraction strategies like using near-far detector joint analysis.} In Fig.~\ref{fig:tree}, we present the 90\% confidence level (C.L.) exclusion contours for the ALP-photon coupling $g_{a\gamma}$ and the ALP-electron coupling $g_{ae}$. With an exposure of 10 kg$\cdot$yr and 2 dru background, RECODE would extend the sensitivity of beam-dump experiments in probing the ALP-photon coupling $g_{a\gamma}$, and is comparable with NEON in the cosmological triangle region. {It is noteworthy that above the 0.2 MeV axion mass, due to the weaker backgrounds in first few bins compared to those in NEON, the exclusion limit do not decline as that of NEON.} While for the ALP-electron coupling $g_{ae}$, RECODE could achieve a similar sensitivity as beam-dump experiments but surpass NEON. We also show the projected limits with an increased exposure of 100 kg·yr and a reduced background of 0.2 dru. It can be seen that RECODE could almost exclude the whole cosmological triangle region for $g_{a\gamma}$, and exceed the sensitivity of current beam-dump experiments for $g_{ae}$.

{\it \textbf{Conclusion}} --
We demonstrate the unique potential of RECODE for MeV-scale ALP searches using intense reactor-produced ALP fluxes. With 10 kg·yr exposure and 2 dru background, our analysis reveals its sensitivity surpassing beam-dump experiments in probing the ALP-photon cosmological triangle and competing with other ALP-electron constraints. Projected improvements (100 kg·yr, 0.2 dru) enable {to nearly cover} the cosmological triangle in the ALP-photon coupling and a new stronger sensitivity of probing ALP-electron coupling. This will significantly expand the reach of terrestrial experiments into previously inaccessible parameter space of axions.

{\it \textbf{Acknowledgments}} -- This work is supported by the National Key Research and Development Program of China (Grant No. 2022YFA1605000). LW is supported by the National Natural Science Foundation of China (NNSFC) under grant No. 12275134 and No. 12335005. YW is supported by the NNSFC under grant No.~12305112. LTY is supported by the NNSFC under grant No. 12322511. Guanhua Gu thanks for the hospitality during his stay in NNU.

\section{Appendix}
\begin{figure}[htbp]
    \centering
    \includegraphics[width=\columnwidth]{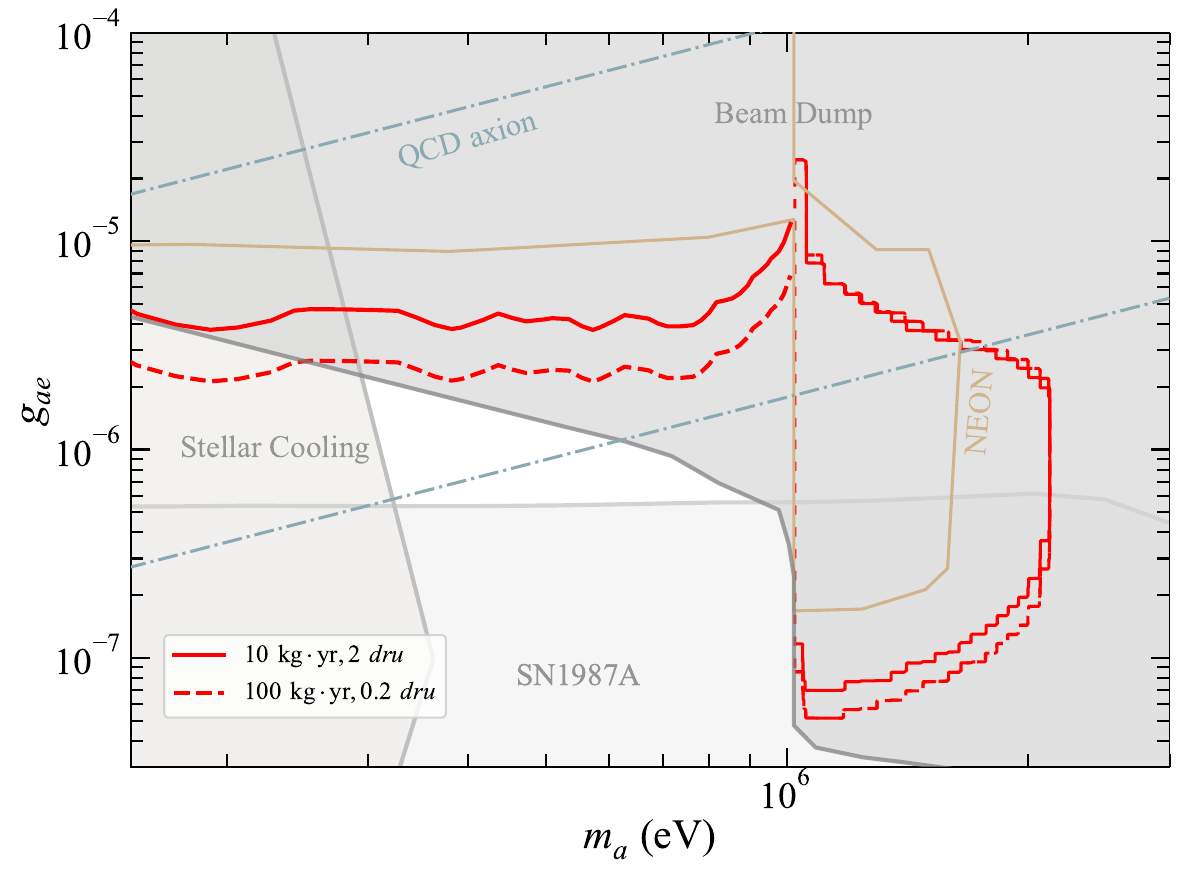}
    \caption{Same as Fig.~\ref{fig:tree}, but for ALP-electron couplings under the assumption that electron-loop mediated decay $a \rightarrow \gamma\gamma$ is not suppressed.}
    \label{fig:loop}
\end{figure}

For ALP-electron coupling, we will have the axion diphoton decay induced by an electron loop as $m_a < 2m_e$. The corresponding decay width is given by,
\begin{equation}
\Gamma_{2\gamma}^{\mathrm{el}}=\frac{g^2_{ae}e^4m_a^3}{256\pi^5m_e^2}f_p({m_a^2}/{4m_e^2}),
\end{equation}
with the loop function, 
\begin{equation}
f_p(x)=\frac{\left|\ln \left[1-2\left(x+\sqrt{x^2-x}\right)\right]^2\right|^2}{64x^2}.
\end{equation}
In the study of ALP-electron coupling, this channel may become dominant at low masses if there is no cancellation from other processes. In Fig.~\ref{fig:loop}, we show the expected constraints on $g_{ae}$ under the assumption that such an loop-induced process is not suppressed. Compared with the the lower panel of Fig.~\ref{fig:tree}, the RECODE sensitivity to probing $g_{ae}$ will be mildly enhanced in the low-ALP-mass range.  

\bibliography{References}

\end{document}